# Xenon Reacts with Iron at the Conditions of the Earth's Core

Li Zhu[1], Hanyu Liu[1], Chris J. Pickard[2], Guangtian Zou[1], and Yanming Ma[1]*

[1]State Key Laboratory of Superhard Materials, Jilin University, Changchun 130012, China

[2]Department of Physics and Astronomy, University College London, Gower Street, London, WC1E 6BT, United Kingdom

**Studies of the Earth's atmosphere have shown that more than 90% of xenon (Xe) is depleted compared with its abundance in chondritic meteorites[1,2]. This long-standing missing Xe paradox has become the subject of considerable interest[3-16] and several models for a Xe reservoir[3-5] have been proposed. Whether the missing Xe is hiding in the Earth's core has remained a long unanswered question[6-9]. The key to address this issue lies in the reactivity of Xe with iron (Fe, the main constituent of the Earth's core), which has been denied by earlier studies[7,9]. Here we report on the first evidence of the chemical reaction of Xe and Fe at the conditions of the Earth's core, predicted through first-principles calculations and unbiased structure searching techniques. We find that Xe and Fe form a stable, inter-metallic compound of $XeFe_3$, adopting a $Cu_3Au$-type face-centered cubic structure above 183 GPa and at 4470 K. As the result of a Xe → Fe charge transfer, Xe loses its chemical inertness by opening up the filled $5p$ electron shell and functioning as a $5p$-like element, whilst Fe is unusually negatively charged, acting as an oxidant rather than a reductant as usual. Our work establishes that the Earth's core is a natural reservoir for Xe storage, and possibly provides the key to unlocking the missing Xe paradox.**

Xe, a noble gas characterized by great chemical inertness, has long been used to study the evolution of Earth and its atmosphere. The unexpected depletion of Xe in the Earth's atmosphere relative to argon (Ar) and krypton (Kr)[1,2], the so-called "missing Xe paradox", has become one of the most challenging enigmas in the planetary sciences. Although models have proposed the possibility of Xe escape from the atmosphere into space[2,10], the majority of researchers accept that Xe may be hidden in the interior of Earth[6-9,11-13].

Attempts towards Xe capture in ices, clathrates, and sediments in crust of Earth have failed[11-13]. A recent experiment reported Xe reactivity with water ice, but the reaction took place at pressure and temperature conditions of 50 GPa and 1500 K found in the interiors of giant planets such as Uranus and Neptune[14]. Studies have also proposed that Xe might be largely retained in the Earth's mantle[3,15,16]. Experimentally, a few percent by weight of Xe has been successfully incorporated into silica at 0.7 GPa and 500 K, where Xe replaced silicon[3]. A subsequent experiment reported the synthesis of Xe oxides[15], whereas a first-principles study demonstrated that incorporation of Xe into silica (by substituting silicon) is energetically unfavorable (the formation energy is greatly positive, at > 3.93 eV per cell) compared to decomposition into silica and Xe[17].

The Earth's core, which contains approximately one third of the Earth's mass, has also been considered as a potential Xe reservoir[7]. However, high-pressure experiments up to 155 GPa have failed to find any evidence of chemical reaction of Xe and Fe[7,9]. A theoretical study proposed that Xe could be trapped in the Earth's core by forming solid solutions with Fe, assuming a hexagonal close-packed lattice[8]. Nishio-Hamane *et al.*[9] soon ruled out this possibility by measuring the compression rates of Xe and Fe, and extrapolating to the pressures relevant to the inner core. They concluded that the atomic sizes of Fe and Xe differ by more than 30%, apparently in contradiction to the Hume-Rothery rules governing the formation of solid solutions.

If Xe were captured in the Earth's core, it would have to form a chemically stable compound with Fe to resist any release into atmosphere. Here, we establish just such a chemical reaction of Xe and Fe on the basis of our structure searching method,[18,19] combined with first-principles calculations.

We first investigated the phase stabilities of various XeFe$_x$ ($x$ = 0.5, 1.0 – 6.0) compounds by calculating their formation enthalpies at 0 K, at pressures of 150, 250 and 350 GPa, relative to the products of dissociation into constituent elements, as summarized in Fig. 1a. The crystal structures used were obtained from our first-principles structure searching simulations and are described in detail below. At 150 GPa, the formation enthalpies of all stoichiometries are positive, which is in consistent with previous reports that Xe and Fe do not react at pressures up to 155 GPa[7,9]. Stable phases emerge at 250 GPa, and XeFe$_3$ is formed as the most stable stoichiometry.

Thermal contributions to the formation free energy are important, since the Earth's core is subject to high temperatures. A recent study extends the temperature of the Earth's inner core beyond 6000 K[20]. We therefore performed quasi-harmonic free-energy calculations with phonon spectra computed using the finite-displacement method (Fig. 1c). We introduced volume dependence of phonon frequencies in an effort to partially account for the anharmonic effect in the free-energy calculations. Thermal effects further stabilize the Xe-Fe compounds by lowering the formation energies of most of the stoichiometries. Still, XeFe$_3$ remains as a thermodynamically most stable species at elevated temperatures. Notably, XeFe$_5$ and XeFe come into play on the convex hull when temperature increases beyond 2000 K (see also supplementary Fig. S1).

Based on a thorough structure search, we find that XeFe$_3$ at 250 GPa and 350 GPa stabilizes into a face-centered cubic Cu$_3$Au-type structure (space group *Pm-3m*, 1 formula unit per cell; see Figs. 2a and 2b). Therefore, every Fe atom has four neighboring Xe and eight neighboring Fe atoms, while every Xe atom is characterized by a 12-fold coordination of Fe atoms by forming XeFe$_{12}$-truncated cubes (Fig. 2b). All 12 nearest Fe-Fe and Fe-Xe distances (2.2 Å at 350 GPa) are equal. XeFe$_5$ exhibits a hexagonal structure (space group *P-62m*, 1 formula unit per cell; see Fig. 2c). The basic building blocks of the structure are also XeFe$_{12}$ polyhedrons. However, the XeFe$_{12}$ polyhedron is a cuboctahedron rather than a truncated cube as in XeFe$_3$. Unlike XeFe$_3$ and XeFe$_5$, XeFe adopts a seven-fold structure (space group *P*-1, 4 formula units per cell; see Fig. 2d) containing a XeFe$_7$ square antiprism composed of face-capped trigonal prisms. Polyhedral views of XeFe reveal an intriguing stacking order of wrinkled Fe-Xe-Fe sandwiches.

Electronic band-structure calculations (Fig. 3a) at 250 GPa revealed the metallic nature of XeFe$_3$, as expected, since at this pressure pure Xe is already a metal. To understand the bonding nature of the hitherto unknown XeFe$_3$ compound, we constructed a model system of the hypothetical XeFe$_0$, in which all Fe atoms were removed from the Cu$_3$Au-type structure, for comparison with the real FeXe$_3$ system at 250 GPa for calculations of projected density of states. Once Fe was incorporated into the lattice, the completely filled 5*p* valence states of Xe were partially depleted to the unoccupied orbital (Fig. 3b). Mulliken's analysis of electron density reveals a transferred charge of approximately 0.72*e* from Xe-5*p* electrons to the Fe-3*d* orbital, which is caused by the Xe-5*p* and Fe-3*d* orbital hybridization. The Mulliken charge result is in good accordance with that (0.75*e*) derived from a direct integration of the projected density of states of the depleted Xe-5*p* in Fig. 3b. This charge-transfer phenomenon is supported further by the difference charge density plot shown in Fig. 3c.

Here, we emphasize the chemical novelty towards to the two end elements, Xe and Fe, during the formation of XeFe$_3$. First, the completely filled 5*p* shell of Xe opens up and, therefore, Xe behaves like a 5*p* element. In view of the very nature of this 5*p*-like Xe, the formation of stable Xe-Fe compounds is not unexpected. Previous studies indeed demonstrated that 5*p* elements (e.g., Te and I) react with Fe[21,22]. Stabilization of this 5*p*-like Xe is not a big surprise, since Xe reacts with oxygen[16] and fluorine[23]. An earlier study even reported a 5*p*-like behavior of Cs in Cs fluoride[24]. However, the peculiarity arises from the reaction of Xe with Fe. Chemically, Fe has an electronegativity of 1.83 on the Pauling scale, apparently lower than that of Xe (2.60), and therefore an electron loss of Fe should be expected by chemical intuition. By contrast, Fe here is unusually negatively charged and turns out to be an oxidant rather than a reductant as it is in most cases. This oxidant nature of Fe might have strong implication on Fe chemistry in the Earth's core. A recent prediction[5] on the chemical interaction of Mg and Xe reveals a fundamentally different bonding characteristic. Xe is negatively charged as an anion, whilst Mg becomes a cation because of its much smaller electronegativity (1.31).

Nickel (Ni) is the second most abundant element in the Earth's core after Fe. Our calculations established that Xe can readily react with Ni under the pressure and temperature conditions of the Earth's core (Fig. 1b and d). The most stable Xe-Ni compound also adopts $XeNi_3$ stoichiometry but the structure is hexagonally packed (space group *Pmmn*, 2 formula units per cell; see Fig. 2e), different with that in $XeFe_3$. Chemically, $XeNi_3$ is rather similar to $XeFe_3$, where Ni is also negatively charged. It appears that Xe reacts more steadily with Ni than with Fe at a substantially lower pressure (120 GPa at 6000 K in Fig. 4). The chemical origin of this reaction might be that without acquiring more electrons from Xe, Ni in its element is already in a higher $d^8$ electron configuration. Indeed, we found a less transferred charge of 0.62$e$ from Xe-5$p$ to Ni-3$d$. We performed extensive structural predictions at 0 K to explore the reaction of Ar and Kr with Fe and Ni, respectively, at pressures up to 400 GPa, but found no sign of reactivity.

Figure 4 shows the computed pressure versus temperature phase diagram of $XeFe_3$ and $XeNi_3$ with respect to the mixture of elemental Xe and Fe or Ni. Our results reveal that $XeFe_3$ and $XeNi_3$ are readily stable under the conditions of the Earth's core (see the isentropic line in Fig. 4). The Earth's core becomes a natural reservoir for depleted Xe. In contrast, because of the nonreactivity of Ar and Kr with Fe and Ni at the conditions of the Earth's core, Ar and Kr have nowhere to hide in the Earth's core; they can only escape into the atmosphere. Notably, the atomic mass of Xe (131.293u) is evidently heavier than that of Fe (55.845u). However, the total mass of the depleted Xe is on the order of $10^{13}$ kg[1], which is about eleven orders of magnitude lower than that of the Earth's core (approximately $1.932 \times 10^{24}$ kg)[25]. Therefore, storage of the entire missing Xe will have a negligible contribution to the total mass and the density of the Earth's core.

We find that the structures of stable $XeFe_3$ and $XeNi_3$ compounds at conditions of the Earth's core are in apparently contrast to the structures of their parent metals (Fe is hexagonally close-packed, while Ni is face-centered cubic). This explains the failure of earlier theoretical calculations[7,8] by assuming that the structures of compounds resemble those in pure metals. Our findings were benefited from the unbiased structure prediction techniques[18,19,26,27] developed recently. Our energetic calculations confirmed that earlier proposed Xe-Fe solid solution model[8] is very unfavorable if compared with our stable $XeFe_3$ compound (Supplementary Table S1). The convex hull calculations (Fig. 1) indicated that besides $XeFe_3/Ni_3$, $XeFe_5/Ni_5$ and $XeNi_6$ can exist in the Fe/Ni-rich regimes (the actual situation in the Earth's core) since they are calculated to sit on the convex hull and are stable with respect to the dissociation into $XeFe_3$ ($XeNi_3$) + Fe (Ni) (see also Supplementary Fig. S1). We do not exclude the existence of other stoichiometries with even higher Fe or Ni contents. Those structure searching calculations are not within the scope of current research. In any regard, our conclusions on the most stable stoichiometry of $XeFe_3/Ni_3$ and the actual reactivity of Fe and Ni with Xe at the conditions of the Earth's core remain unaltered.

The case of Xe-Fe chemical reaction under the conditions of the Earth's core not only helps to uncover the missing Xe paradox, but also provides unexpected novel chemistry of Xe and Fe created by compression of two seemingly incompatible elements. The generation of negatively charged Fe and the $5p$-like Xe makes the reaction possible. The present findings might also shed light on the Earth's evolution by virtue of the model of the missing Xe in the Earth's core.

**METHODS SUMMARY**

Our structure searching simulations are performed through the calypso method[18] based on a global minimization of free energy surfaces merging *ab initio* total-energy calculations as implemented in the CALYPSO code[19], which is specially designed for global structural minimization unbiased by any known structural information. Our method has been benchmarked on various known systems, ranging from elements to binary and ternary compounds[18]. The results of the structural searches were subsequently confirmed using the Ab Initio Random Structure Searching approach to structure prediction[26,27]. Total energy calculations were carried out with the CASTEP plane-wave code[28], ultrasoft pseudopotentials and the Perdew-Burke-Ernzerhof generalized gradient approximation functional. The pseudopotentials used treat $4d^{10}5s^25p^6$, $3s^23p^63d^64s^2$ and $3s^23p^63d^84s^2$ as valence electrons for Xe, Fe and Ni atoms, respectively. The use of the plane-wave kinetic energy cutoff of 600 eV and dense $k$-point sampling adopted here, were shown to give excellent convergence of total energies. We explored the effects of temperature using the quasiharmonic approximation that introduces volume dependence of phonon frequencies as a part of anharmonic effect, for which phonon calculations were performed for all promising structures using the phonopy code[29]. In order to ensure the validity of the pseudopotentials adopted here, we have also performed full-potential all-electron calculations for the equation of states of XeFe$_3$ and reaction pressures of Xe and Fe by using the WIEN2K code[30]. Our all-electron calculations gave nearly identical results as CASTEP calculations did, validating our pseudopotentials.

**Acknowledgements** We thank the China 973 Program (2011CB808200), Natural Science Foundation of China under Nos. 11274136, 11104104, 11025418 and 91022029, the 2012 Changjiang Scholars Program of China, Changjiang Scholar and Innovative Research Team in University (IRT1132). Part of the calculations was performed in the High Performance Computing Center of Jilin University. C. J. P. was funded by the United Kingdom's EPSRC.

**Author Contributions** Y.M. proposed the research. L.Z., H.L., C. J. P., and Y.M. did the calculations. L.Z., G.Z., C. J. P., and Y.M. analysed the data. L.Z. and Y.M. wrote the paper. All authors read and commented on the manuscript.

**Author Information** Correspondence and request for materials should be addressed to Y.M. (mym@jlu.edu.cn).

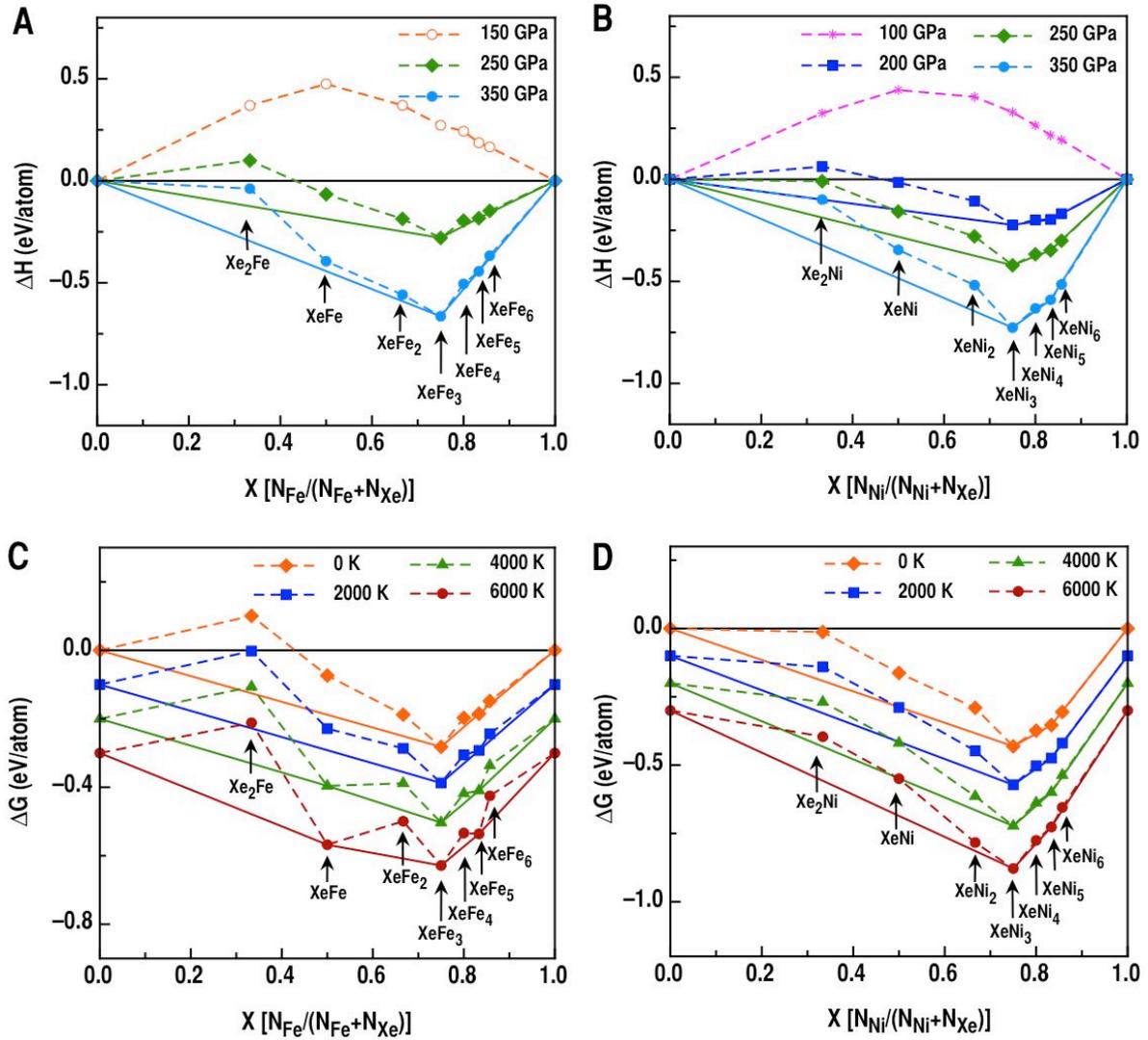

**Fig. 1. Phase stabilities of Xe-Fe and Xe-Ni compounds**. (**A-B**) Predicted formation enthalpy of various Xe-Fe (A) and Xe-Ni (B) compounds with respect to the elemental decomposition into Xe and Fe at 0 K and high pressures. (**C-D**) Predicted Gibbs free energy of various Xe-Fe (C) and Xe-Ni (D) compounds at 250 GPa relative to the elemental decomposition into Xe and Fe as a function of temperatures. For clarity, we downshift the temperature dependent convex hull by a constant of 0.1 eV per 2000 K interval.

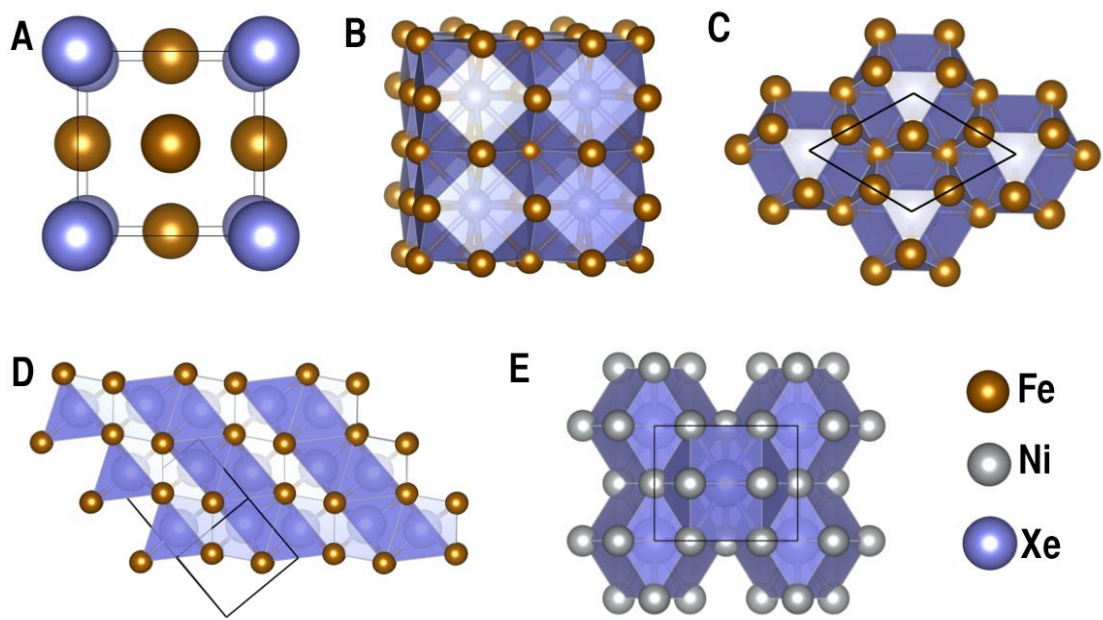

**Fig. 2. Selected structures of Xe-Fe compounds. (A)** Top view of Cu$_3$Au-type structure of XeFe$_3$. **(B-D)**, Polyhedral views of Cu$_3$Au-type structure of XeFe$_3$ (B), *P*-62*m* structure of XeFe$_5$ (C), and *P*-1 structure of XeFe (D).

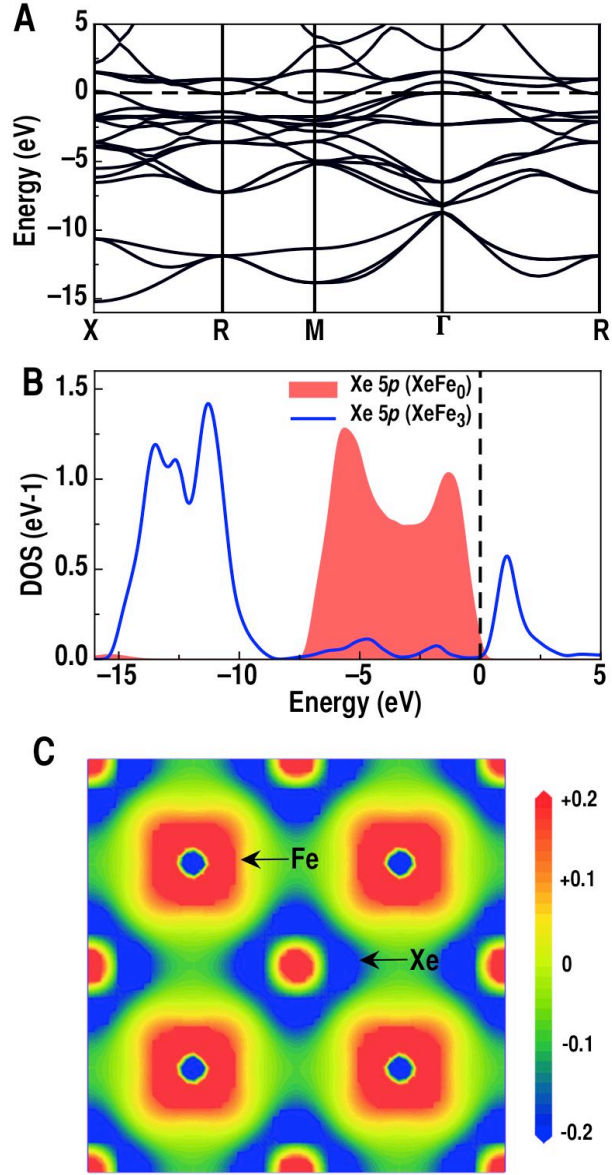

**Fig. 3. Electronic properties of XeFe$_3$.** **(A)** The electronic band structure for XeFe$_3$ at 250 GPa. The dashed line indicates the Fermi energy of XeFe$_3$. **(B)** Projected densities of states of Xe-5*p* states for XeFe$_3$ and hypothetical XeFe$_0$ at 250 GPa. The dashed line indicates the Fermi energy. **(C)** The difference charge density (eÅ$^{-3}$; crystal density minus superposition of isolated atomic densities) of XeFe$_3$ plotted in the (100) plane at 250 GPa.

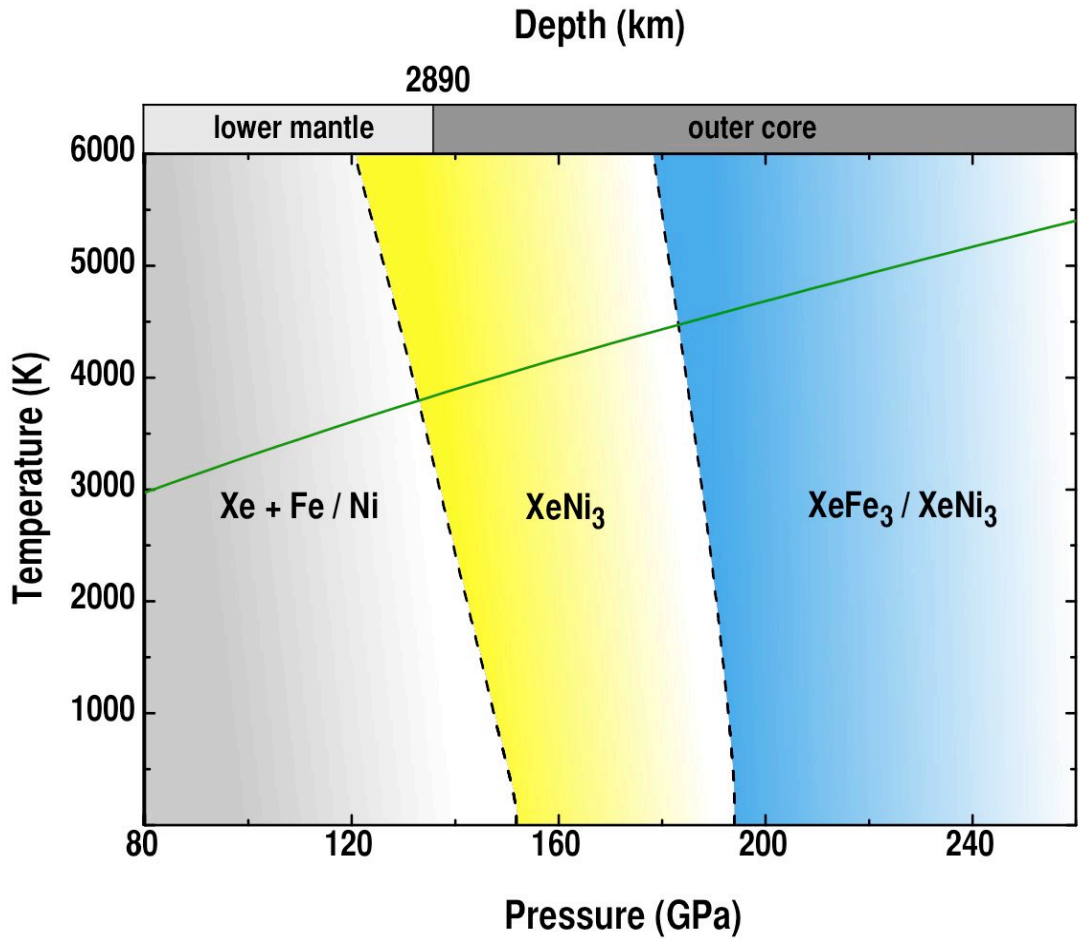

**Fig. 4. Phase diagrams of Xe-Fe and Xe-Ni systems**. The dashed lines show the proposed phase boundaries and the green solid line presents the isentrope of the Earth's core from Ref. 20. The left and gray colored region is the mixture of elemental Xe with Ni and Fe. The middle and yellow colored region depicts the formation diagram of stable $XeNi_3$, but not $XeFe_3$. The right and blue colored region illustrates the phase diagram for the formation of both $XeNi_3$ and $XeFe_3$.